\documentclass[conference]{IEEEtran}
\IEEEoverridecommandlockouts
\usepackage{cite}
\usepackage{amsmath,amssymb,amsfonts}
\usepackage{algorithmic}
\usepackage{graphicx}
\graphicspath{{figures/}}
\usepackage{textcomp}
\usepackage{xcolor}
\usepackage{hyperref} 
\usepackage{subfigure} 
\usepackage{bm}
\def\BibTeX{{\rm B\kern-.05em{\sc i\kern-.025em b}\kern-.08em
		T\kern-.1667em\lower.7ex\hbox{E}\kern-.125emX}}

\begin{document}
\makeatletter
\newcommand{\linebreakand}{%
  \end{@IEEEauthorhalign}
  \hfill\mbox{}\par
  \mbox{}\hfill\begin{@IEEEauthorhalign}
}
\makeatother

\title{Rethinking the Role of Operating Conditions for Learning-based Multi-condition Fault Diagnosis}

\author{
	\IEEEauthorblockN{Pengyu Han}
	\IEEEauthorblockA{\textit{Department of Automation} \\
		\textit{Tsinghua University}\\
		Beijing, China \\
		hpy24@mails.tsinghua.edu.cn}
	\and
	\IEEEauthorblockN{Zeyi Liu}
	\IEEEauthorblockA{\textit{Department of Automation} \\
		\textit{Tsinghua University}\\
		Beijing, China \\
		liuzy21@mails.tsinghua.edu.cn}
	\and
	\IEEEauthorblockN{Shijin Chen}
	\IEEEauthorblockA{\textit{MCC5 Group} \\
	    \textit{Shanghai Co. LTD}\\
		Shanghai, China \\
		chenshijin1993@gmail.com
	}
    \linebreakand
	\IEEEauthorblockN{Dongliang Zou}
	\IEEEauthorblockA{\textit{MCC5 Group} \\
	    \textit{Shanghai Co. LTD}\\
		Shanghai, China \\
		DLZOU2006@163.com
	}
	\and
	\IEEEauthorblockN{Xiao He}
	\IEEEauthorblockA{\textit{Department of Automation} \\
		\textit{Tsinghua University}\\
		Beijing, China \\
		hexiao@tsinghua.edu.cn
	}
	
	\thanks{This work was supported by National Natural Science Foundation of China under grant 62473223 and 624B2087, and Beijing Natural Science Foundation under grant L241016. (\emph{Corresponding author: Xiao He.})
	}
}

\maketitle
\begin{abstract}
Multi-condition fault diagnosis is prevalent in industrial systems and presents substantial challenges for conventional diagnostic approaches.
The discrepancy in data distributions across different operating conditions degrades model performance when a model trained under one condition is applied to others.
With the recent advancements in deep learning, transfer learning has been introduced to the fault diagnosis field as a paradigm for addressing multi-condition fault diagnosis.
Among these methods, domain generalization approaches can handle complex scenarios by extracting condition-invariant fault features.
Although many studies have considered fault diagnosis in specific multi-condition scenarios, the extent to which operating conditions affect fault information has been scarcely studied, which is crucial.
However, the extent to which operating conditions affect fault information has been scarcely studied, which is crucial.
When operating conditions have a significant impact on fault features, directly applying domain generalization methods may lead the model to learn condition-specific information, thereby reducing its overall generalization ability.
This paper investigates the performance of existing end-to-end domain generalization methods under varying conditions, specifically in variable-speed and variable-load scenarios, using multiple experiments on a real-world gearbox.
Additionally, a two-stage diagnostic framework is proposed, aiming to improve fault diagnosis performance under scenarios with significant operating condition impacts.
By incorporating a domain-generalized encoder with a retraining strategy, the framework is able to extract condition-invariant fault features while simultaneously alleviating potential overfitting to the source domain.
Several experiments on a real-world gearbox dataset are conducted to validate the effectiveness of the proposed approach.

\end{abstract}

\begin{IEEEkeywords}
	Fault diagnosis, multi-condition, domain generalization, random vector functional link
\end{IEEEkeywords}


\section{Introduction}
During industrial production processes, system operating conditions are influenced by various factors such as changes in production demands and environmental disturbances, exhibiting complex and dynamic multi-condition characteristics that increase the difficulty of fault diagnosis \cite{Song2024,li2025dynamic,Chen2023a,Liu2023a}. 
Although machine learning-based methods have received widespread attention in the field of fault diagnosis in recent years, most approaches still assume that the training and test sets satisfy the independent and \emph{identically distributed} (i.i.d.) condition \cite{Wang2021,Fan2024a}.
However, in practical industrial applications, the continuous variation of operating conditions leads to significant differences in data distributions across different conditions, making the i.i.d. assumption difficult to hold \cite{Tang2024,Yao2023,Lu2023b}. 
The distribution shift causes traditional machine learning-based fault diagnosis methods to suffer from performance degradation under multi-condition environments. 
As a result, the model's generalization ability is constrained, making it challenging to meet the demands of complex and dynamic real-world scenarios.
Therefore, research on intelligent fault diagnosis under multi-condition scenarios holds not only important theoretical significance but also broad practical application value \cite{Li2024a,Hei2024,10530529,10297982}.

To address the distribution discrepancies caused by multi-condition scenarios, transfer learning has been introduced as an effective learning paradigm in the field of multi-condition fault diagnosis \cite{han2024multi,Chen2024,Yang2023a,Li2023}. 
Specifically, domain adaptation-based methods have been extensively studied \cite{Yang2024,Wu2024,YU2022111597,Zhang2022}. These methods typically transfer knowledge from the source domain to the target domain through data alignment techniques.
Yang \emph{et al.} proposed a deep adversarial hybrid domain adaptation network (DAHAN), which leverages local domain alignment using the Wasserstein distance to match class-level features \cite{Yang2023a}. Zhang \emph{et al.} introduced a maximum mean discrepancy loss to achieve cross-domain alignment within modalities, demonstrating accurate fault diagnosis performance on a variable-speed gearbox \cite{Zhang2024}. 
However, these methods rely on the assumption that a certain amount of target domain data is available.
In practical scenarios, only limited stable condition data can be collected offline. When encountering unseen conditions during online operation, condition identification cannot be performed in real-time, making the target domain inaccessible.
In this context, domain generalization has emerged as a promising paradigm to tackle such challenge \cite{Liu2024a,Zheng2023,Chen2023c,Wang2024a}. 
These approaches typically learn domain-invariant features across multiple source domains to achieve robust generalization to unseen operating conditions. 
In \cite{Wang2023}, an adaptive class center generalization network (ACCGN) was proposed, which employs adaptive center loss and a sparse domain regression framework to learn class-level and domain-level invariant representations. Yu \emph{et al.} introduced a framework combining graph neural networks (GNNs) with a dynamic graph embedding mechanism, achieving stable and accurate diagnostic performance under variable operating conditions \cite{Yu2023a}. 

However, a critical issue remains: \emph{to what extent do operating conditions affect fault information?}
Taking vibration signal analysis in mechanical systems as an example, variations in rotational speed and load have markedly different effects on signal characteristics. Rotational speed changes are generally considered to induce stronger condition variations than load changes, as speed variations directly alter the excitation frequency components and energy distribution of the system, significantly affecting the frequency-domain characteristics of vibration signals. 
In contrast, load changes primarily influence the amplitude of the signal, with relatively minor impact on its frequency structure. 
The strong condition features introduced by rotational speed variations can obscure underlying fault patterns, making it more challenging for transfer learning-based fault diagnosis models to accurately extract consistent fault representations across conditions. 
Moreover, directly applying domain generalization methods to multiple source domains with large condition discrepancies may lead the model to learn features entangled with condition-specific information, thereby reducing the generalization ability of model.

\color{black}
To validate the impact of operating condition information on fault features, multiple experiments are conducted to analyze the performance differences of the existing domain generalization end-to-end method under varying conditions, including variable-speed and variable-load scenarios.
Additionally, a two-stage diagnostic framework is introduced in this paper to improve fault diagnosis performance in scenarios where the impact of operating conditions is significant.
The method first employs a domain generalization approach to learn condition-invariant fault features from multiple source domains.
A retraining strategy is then utilized to alleviate potential overfitting to source domains, thereby enhancing overall generalization and robustness.

\color{black}

The rest of the paper is organized as follows:
In Section II, the problem statement is summarized.
The proposed framework is described in detail in Section III.
Section IV provides a detailed introduction to the real gearbox dataset and experimental setup, and offers an in-depth discussion and analysis of the experimental results.
Finally, the conclusion of this paper is given in Section V.
\color{black}
\section{Problem Statement}

Consider a real-time system that generates a data stream \(\{(\bm x_1, y_1), (\bm x_2, y_2), \dots, (\bm x_t, y_t)\}\), where at each time step \(t\), a sample \(\bm x_t \in \mathbb{R}^{1 \times k}\) consisting of readings from \(k\) sensors is observed, and the corresponding label is denoted as \(y_t\).  

During the offline phase, data can be collected from \(M\) stable operating conditions, denoted as \(\{S_1, S_2, \dots, S_M\}\), where each stable condition contains samples corresponding to the same \(T\) fault types. 

In the online deployment phase, due to variations in equipment load, switching of operational modes, and other dynamic factors, the system often transitions between multiple steady conditions. This inevitably results in the occurrence of transitional conditions. Compared with stable conditions, transitional conditions are influenced by multiple complex factors, such as switching speed, the degree of discrepancy between conditions, and control strategies. As a result, their data distributions typically exhibit strong uncertainty and temporal variability.  
Since transitional conditions are difficult to collect in practical industrial processes, such data are typically considered absent from the offline training set.

\color{black}
\section{Methodology}
\subsection{Case 1: End-to-End Diagnostic Framework}

To enhance the generalization capability of the model under varying working conditions, this paper introduces a classical domain generalization method, named \emph{meta learning for domain generalization} (MLDG) \cite{MLDG}. 
This method simulates distribution shifts between training and testing data by partitioning multiple source domains into meta-train and meta-test sets, enabling the model to acquire cross-domain generalization capability during training.
Assuming that there are $N$ source domains in the training set, each corresponding to a different operating condition. 
Let $\mathcal{F}(\cdot)$ denote the feature extractor, termed the domain-generalized encoder (DGE), and $\mathcal{C}(\cdot)$ the classifier. The overall model can thus be formulated as $\mathcal{C}(\mathcal{F}(\cdot))$.

During training, $P$ domains are selected from the $N$ source domains to form the meta-train set, while the remaining $N - P$ domains serve as the meta-test set. The optimization objective of MLDG is to update the model parameters $\Theta$ on the meta-train set and evaluate the generalization performance on the meta-test set. The objective function is defined as:
\begin{equation}
	\min_{\Theta} \mathcal{L}_{\text{mtrain}}(\Theta) + \gamma \mathcal{L}_{\text{mtest}}(\Theta - \alpha \nabla_{\Theta} \mathcal{L}_{\text{mtrain}}(\Theta)),
\end{equation}
where $\alpha$ denotes the step size for updating on the meta-train set, and $\gamma$ denotes the weight for the meta-test loss.
$\mathcal{L}_{\text{mtrain}}$ and $\mathcal{L}_{\text{mtest}}$ represent the weighted average losses on the meta-train and meta-test sets, respectively. Both loss terms are computed using the cross-entropy loss, defined as:
\begin{equation}
	\mathcal{L}_{\text{mtrain}} = \frac{1}{P} \sum_{i=1}^{P} \frac{1}{n_i}\sum_{j=1}^{n_i} \ell( y_{i}^{j},\hat{y}_{i}^j),
\end{equation}
and
\begin{equation}
	\mathcal{L}_{\text{mtest}} = \frac{1}{N-P} \sum_{i=1}^{N-P}\frac{1}{n_i} \sum_{j=1}^{n_i} \ell( y_{i}^{j},\hat{y}_{i}^j),
\end{equation}
where $n_i$ denotes the number of samples in the $i$-th source domain.
$y_{i}^{j}$ and $\hat{y}_{i}^{j}$ denote the true and predicted labels of the $j$-th sample in the $i$-th source domain, respectively.
$\ell(y, \hat{y}) = -\hat{y} \log(y)$ represents the cross-entropy between  the true label $y$ and the predicted label $\hat{y}$.

In the proposed framework, the DGE $\mathcal{F}$ is implemented using a \emph{multi-scale convolutional neural network} (MSCNN), and the classifier $\mathcal{C}$ is a \emph{multilayer perceptron} (MLP). A schematic illustration of the model structure is shown in Fig.~\ref{MSCNN}.

\begin{figure}[htbp]
	\centering
	\includegraphics[width=0.48\textwidth]{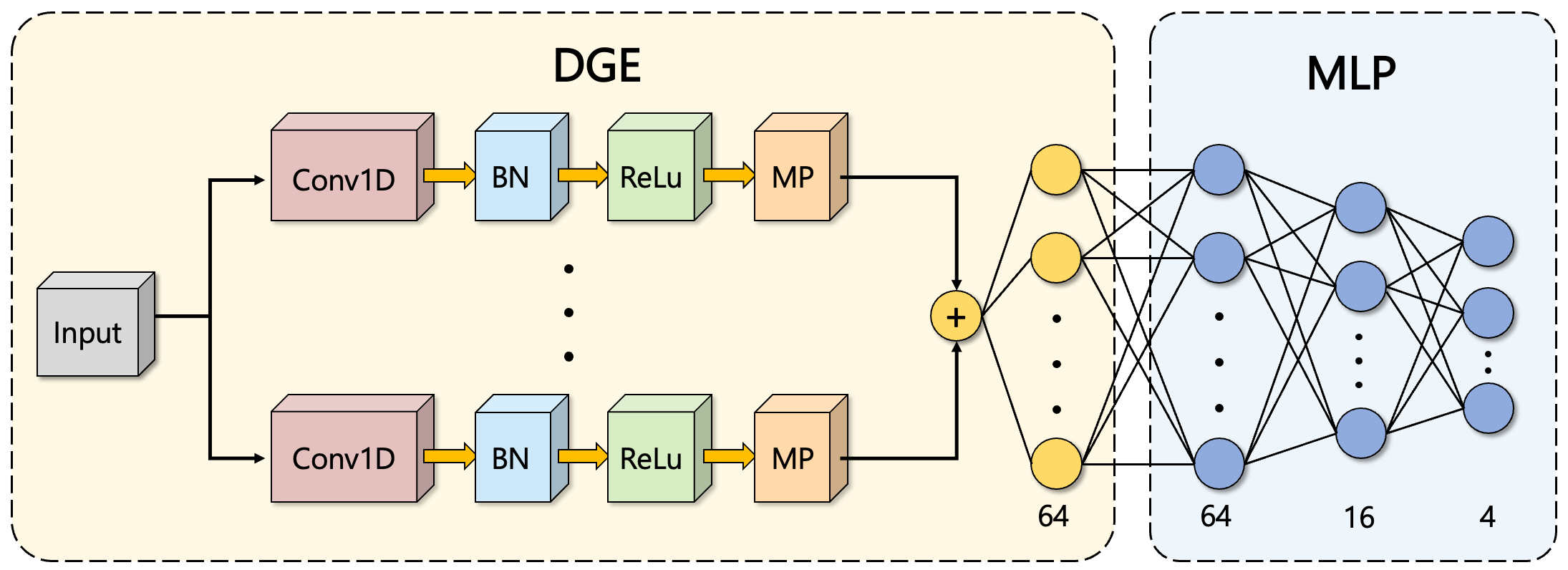}
	\caption{Schematic of the end-to-end diagnostic framework.}
	\label{MSCNN}
\end{figure}

\subsection{Case 2: Two-stage Diagnostic Framework}

To construct a unified embedding space, the proposed framework removes the MLP classifier from the original structure, retaining only MSCNN as a fixed DGE for both source and target domains. 
For any input signal $\bm x$, the corresponding feature representation is obtained as $\bm z = \mathcal{F}(\bm x)$.
Since the output layer of MSCNN is a fully connected layer with 64 neurons, the extracted feature $\bm z$ lies in $\mathbb{R}^{64}$.
All source domain samples in the training set are reprocessed through the fixed DGE $\mathcal{F}$ to obtain feature representations $\bm Z_1, \bm Z_2, \dots, \bm Z_N$, which are mapped into a shared feature space.
A new classifier is then trained on these representations for fault diagnosis.
In this work, a \emph{random vector functional link} (RVFL) network \cite{RVFL} is adopted, which is a single-layer feedforward neural network characterized by its simple architecture and efficient training. 
Figure~\ref{DGE+RVFL} presents a schematic diagram of the proposed framework.

The RVFL network consists of an input layer, a hidden layer, and an output layer.
Specifically, the hidden layer weights $\bm w$ and biases $\bm b$ are randomly initialized and fixed during training.
The output weights $\bm \beta$ are computed analytically, eliminating the need for iterative backpropagation.

The RVFL consists of an input layer, a hidden layer, and an output layer. Let the input layer have $J$ neurons, the hidden layer $Q$ neurons, and the output layer $V$ neurons. For input features $\bm Z$ the hidden layer output $\bm H$ is computed as:

\begin{equation}
	\bm H(\bm Z) = \left[ \bm {h}_1(\bm {Z}),\bm {h}_2(\bm {Z}), \dots, \bm {h}_Q(\bm {Z})\right],
\end{equation}
and
\begin{equation}
	\bm h_{i}(\bm Z) = g\left(\bm w_i \bm Z + \bm b_i \right),
\end{equation}
where $\bm{h}_{i}(\bm{Z})$ denotes the output of the $i$-th hidden neuron, and $g(\cdot)$ represents a nonlinear activation function, typically selected as the sigmoid function.

By concatenating the input features and hidden outputs, the output matrix $\bm{E}$ can be obtained, as shown below:
\begin{equation}
	\begin{aligned}
		\bm{E} 	&= \left[\begin{matrix}\bm{Z} \mid \bm{H}\end{matrix}\right]\\
		&= \left[\begin{array}{ccc|ccc}
			z_{11} & \dots & z_{1J} & \bm h_{1}(\bm z_1) & \dots & \bm h_{Q}(\bm z_1) \\
			\vdots & \ddots & \vdots & \vdots & \ddots & \vdots \\
			z_{N1} & \dots & z_{NJ} & \bm h_{1}(\bm z_N) & \dots & \bm h_{Q}(\bm z_N)
		\end{array}\right].
	\end{aligned}
\end{equation}

Accordingly, the predicted output of the RVFL model $\bm{\hat{Y}}$ can be obtained as follows:
\begin{equation}
	\bm \hat{Y} = \bm E \bm \beta,
\end{equation}
where $\bm \beta = \left[\bm \beta_1, \bm \beta_2, \dots, \bm \beta_{J+Q} \right]^\top $ is the output weight matrix to be learned. The objective is to minimize the $\mathcal{L}_2$ norm between the predicted labels $ \hat{\bm Y}$ and the ground-truth labels $\bm Y$:

\begin{equation}
	\min_{\bm \beta} \bigl\| \bm Y - \bm E \bm \beta\ \bigr\|^2 + \sigma  \bigl\| \bm \beta \bigr\|^2,
\end{equation}
where $\sigma$ is a regularization parameter to prevent overfitting by penalizing overly complex models. By incorporating an $\mathcal{L}_2$ regularization term into the objective function, the model is prevented from relying excessively on individual features, thereby mitigating the risk of overfitting.  
The analytical solution for $\bm{\beta}$ can be obtained by solving the objective function as follows:
\begin{equation}
	\bm \beta = (\bm E^\top \bm E + \sigma \bm I)^{-1} \bm E^\top \bm Y.
\end{equation}

\color{black}

\begin{figure}[htbp]
	\centering
	\includegraphics[width=0.48\textwidth]{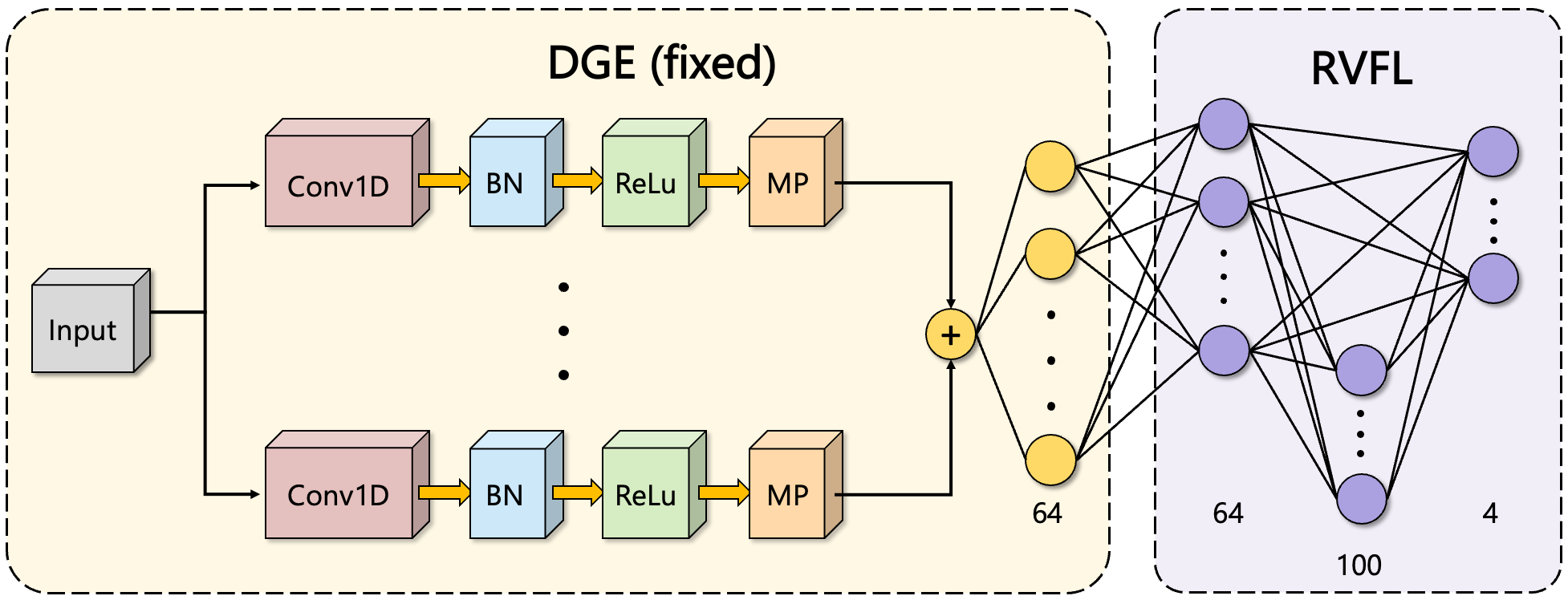}
	\caption{Schematic of the two-stage diagnostic framework.}
	\label{DGE+RVFL}
\end{figure}

\section{Experiment}
\subsection{Experimental Setup}

In this section, the multi-mode gearbox fault diagnosis tasks under time-varying working conditions are considered.  
In this study, the MCC5-THU dataset is used. The experimental setup is illustrated in Fig.~\ref{testrig}, which consists of a 2.2 kW three-phase asynchronous motor, a torque sensor, a two-stage parallel gearbox, a magnetic powder brake acting as a torque generator, and a measurement and control system.  
\begin{figure}[!htbp]
	\centering
	\includegraphics[width=0.47\textwidth]{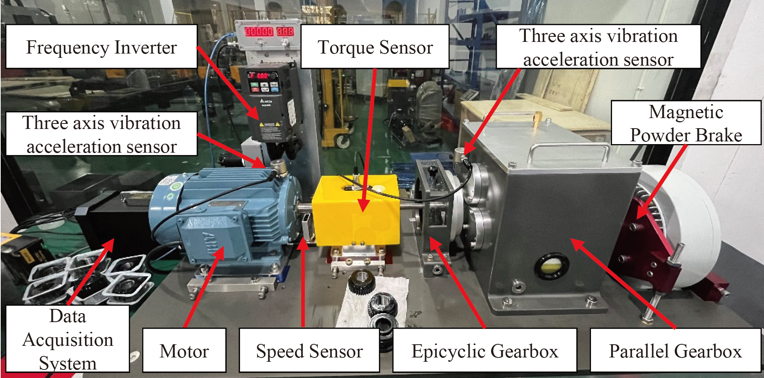}
	\caption{The actual gearbox test rig.}
	\label{testrig}
\end{figure}

The dataset contains data for various fault types and severities under varying rotational speeds and torques. Each sampling session lasts 60 seconds with a sampling frequency of 12.8~kHz, resulting in a total of 240 csv files. 
Each file contains eight-dimensional signals, including speed, load, and two sets of x, y, z-axis vibration signals.  
Since operating condition information is typically unavailable in real-world scenarios, only the six-dimensional vibration signals are used as raw input in subsequent experiments.  
For more details on the dataset, please refer to \cite{MCC5THU}.

During the subsequent model training process, the 6-dimensional raw signals are segmented using a sliding window with a length of 1024 and a step size of 64. The resulting segments are then fed into the model in batches of size 64, producing an input tensor of dimension $\mathbb{R}^{64 \times 6144}$ for each iteration.

\subsection{Comparison Study}

\begin{figure}[!htbp]
	\centering
	\subfigure[Time-varying speed, gear wear]{
		\includegraphics[width=0.48\textwidth]{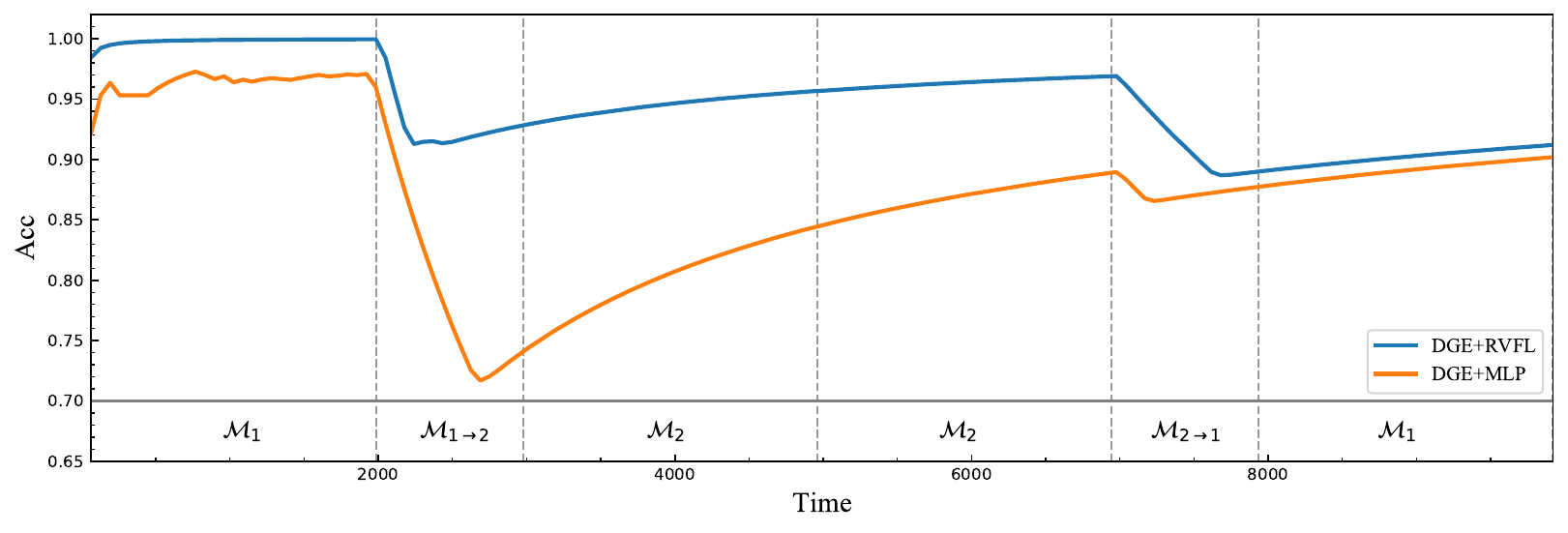}
	}
	\centering
	\subfigure[Time-varying speed, teeth break]{
		\includegraphics[width=0.48\textwidth]{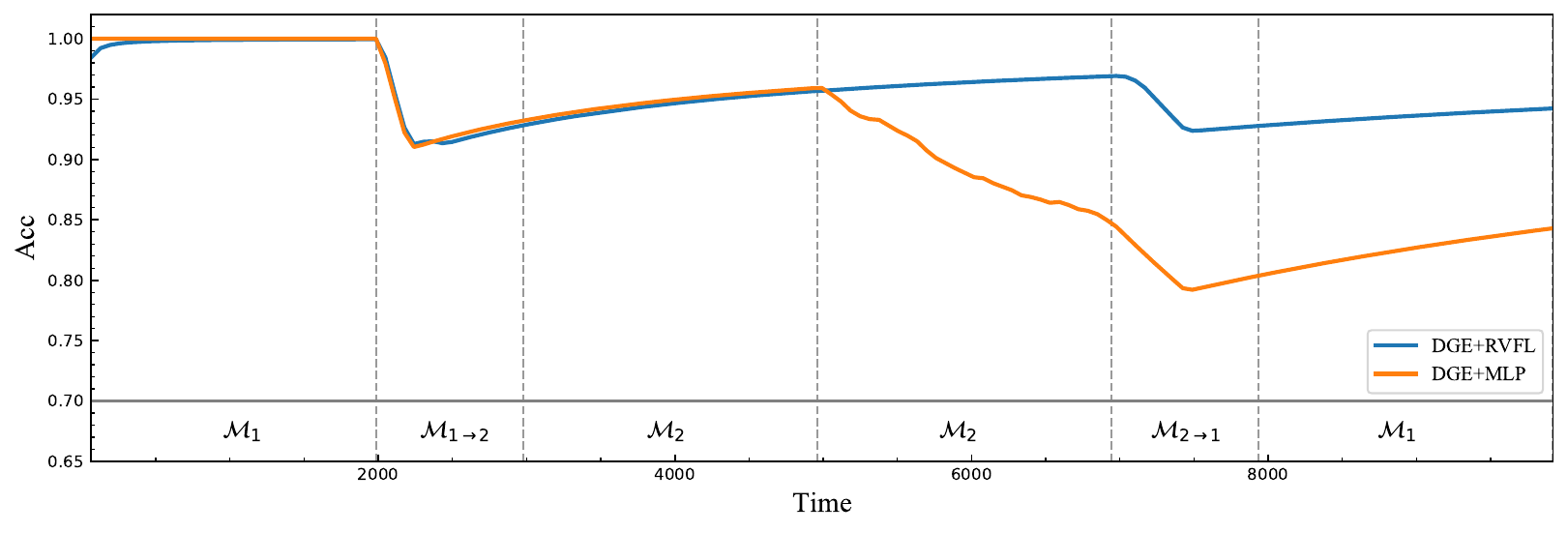}
	}
	\centering
	\subfigure[Time-varying speed, teeth crack]{
		\includegraphics[width=0.48\textwidth]{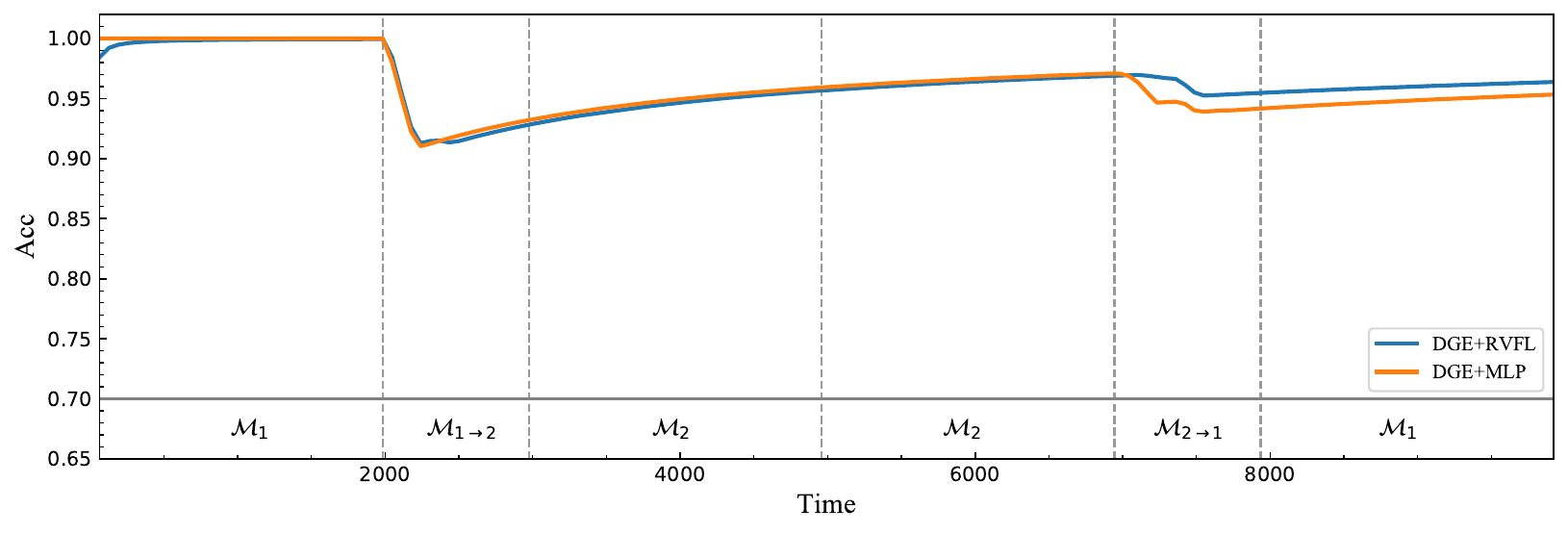}
	}
	\caption{The results of learning curves in the comparison study.}
	\label{CS}
\end{figure}

\begin{figure*}[htbp]
	\centering
	\subfigure[Time-varying speed]{
		\includegraphics[width=0.48\textwidth]{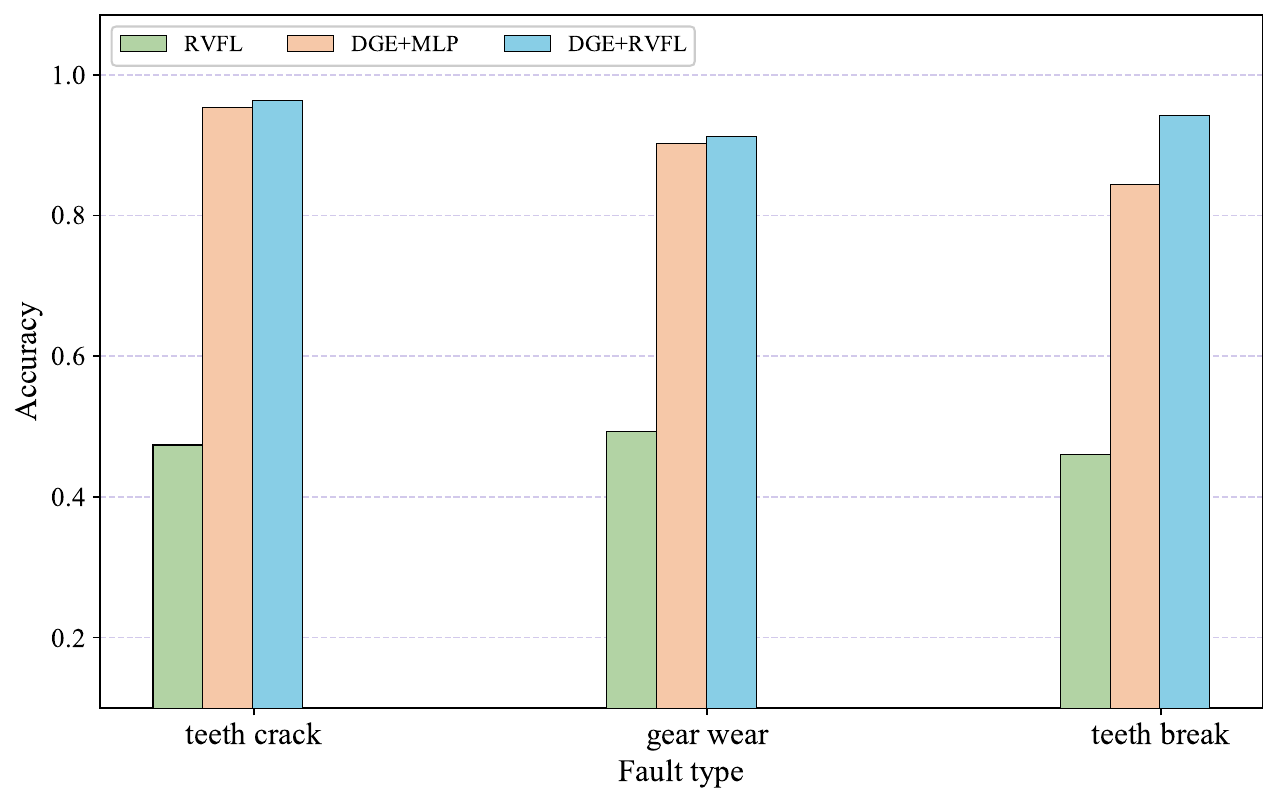}
        \label{bar_speed}
	}
	\centering
	\subfigure[Time-varying torque]{
		\includegraphics[width=0.48\textwidth]{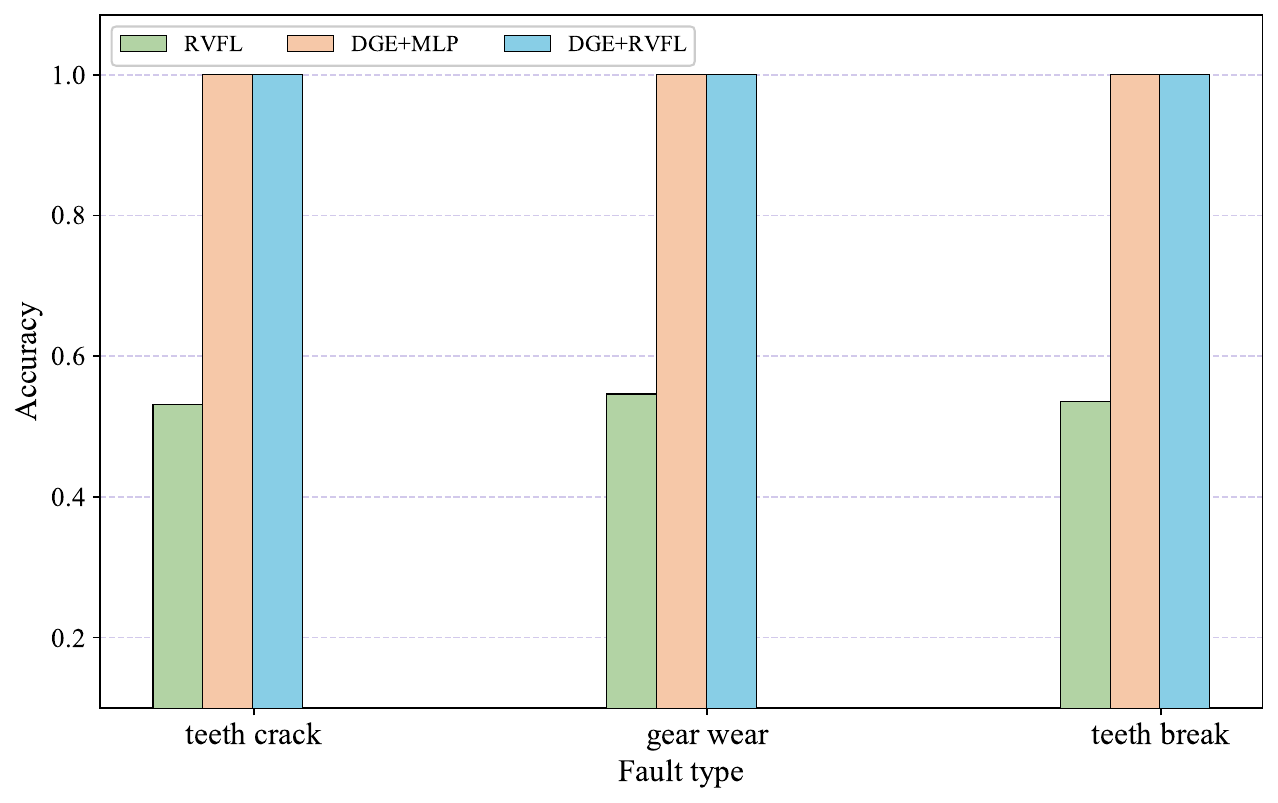}
        \label{bar_torque}
	}
	\caption{The bar plot results for comparison study.}
	\label{bar}
\end{figure*}

Two sets of experiments were conducted to simulate online scenarios with time-varying speeds and loads. 
These experiments were designed to reflect realistic industrial settings, where the operating condition transitions from $\mathcal{M}_1$ to $\mathcal{M}_2$ and then back to $\mathcal{M}_1$. 
At the 4984-th sample point under condition $\mathcal{M}_2$, a specific fault type was introduced.
Each set of experiment shares an identical offline training dataset configuration, encompassing two steady operating conditions (M = 2), each containing one healthy state and three distinct fault categories (N = 4). For each state, 1984 samples were collected, resulting in a total of 15872 samples in the offline dataset. 
It is important to note that although only one fault type appears in each online experiment, the objective is to identify the specific fault type. 
Therefore, the task remains a fault diagnosis problem rather than simple fault detection.

For the variable-speed experimental setup, the load was fixed at 20 Nm, with the rotational speed set to 2000 rpm under condition $\mathcal{M}_1$ and 1500 rpm under condition $\mathcal{M}_2$.  
For the variable-load experimental setup, the speed was fixed at 1000 rpm, with the load set to 20 Nm under condition $\mathcal{M}_1$ and 15 Nm under condition $\mathcal{M}_2$.
The experiment compared the fault diagnosis performance of a direct end-to-end approach (i.e., using MLP for classification), the RVFL model trained directly on raw signals, and the proposed framework.
In the subsequent experiments, the RVFL network was configured with 100 hidden neurons and a regularization coefficient of $\sigma = 0.0001$. 
In MLDG training, the step size is set to $\alpha = 0.005$. The meta-test loss weight $\gamma$ is progressively increased during training using an exponential schedule to enhance stability in the early training stages.

The performance evaluation metric used in this study is cumulative accuracy, with the experimental results shown in Fig.~\ref{CS} and Fig.~\ref{bar}. 
In practical applications, fluctuations in rotational speed exert a greater influence on vibration signals than variations in load. 
Consequently, operating condition information is considered to have a more significant effect in variable-speed tasks, whereas its impact is comparatively limited in variable-load tasks.

It is observed that, for the variable-speed scenario, the proposed algorithm demonstrates a significant advantage over traditional end-to-end methods. 
Specifically, for the diagnosis of {\it teeth break} fault, the RVFL model achieves an average accuracy of 0.95, whereas the MLP model only reaches 0.80. 
This discrepancy may be attributed to potential overfitting when directly employing the end-to-end DGE+MLP model.
In the MLDG framework, although the loss function is designed to improve cross-domain generalization, the training of the classifier $\mathcal{C}$ aims to minimize classification errors in the source domain. 
This can inadvertently lead the model to learn discriminative features that are highly coupled with the specific operating conditions. 
Such coupling is effective in the source domain but limits generalization when the model encounters data with varying conditions in the test set.
In contrast, decoupling the MSCNN and classifier, and retraining the features output by the MSCNN with RVFL, effectively alleviates the protential overfitting problem.

However, for the variable-torque task, both the end-to-end approach and the proposed method achieved 100\% accuracy.
This indicates that when condition information is weak, the end-to-end model is capable of effectively extracting fault features that are invariant across multiple stable conditions, without the need for retraining using the RVFL classifier.
Additionally, as shown in Figs.~\ref{bar_speed} and \ref{bar_torque}, the RVFL model trained directly on raw signals performs poorly in fault diagnosis. This further demonstrates the effectiveness of the DGE in extracting domain-invariant features.

\color{black}
\subsection{Feature Visualization}

In this section, the t-SNE method is applied to reduce the dimensionality of both the original six-dimensional vibration signals and the extracted features obtained from the domain-generalized encoder to two dimensions for visualization. The visualization results under variable-speed and variable-torque scenarios are shown in Figs.~\ref{raw_speed}-\ref{mapped_torque}, respectively.

\begin{figure}[!htbp]
	\centering
	\subfigure[Time-varying speed, raw signals]{
		\includegraphics[width=0.22\textwidth]{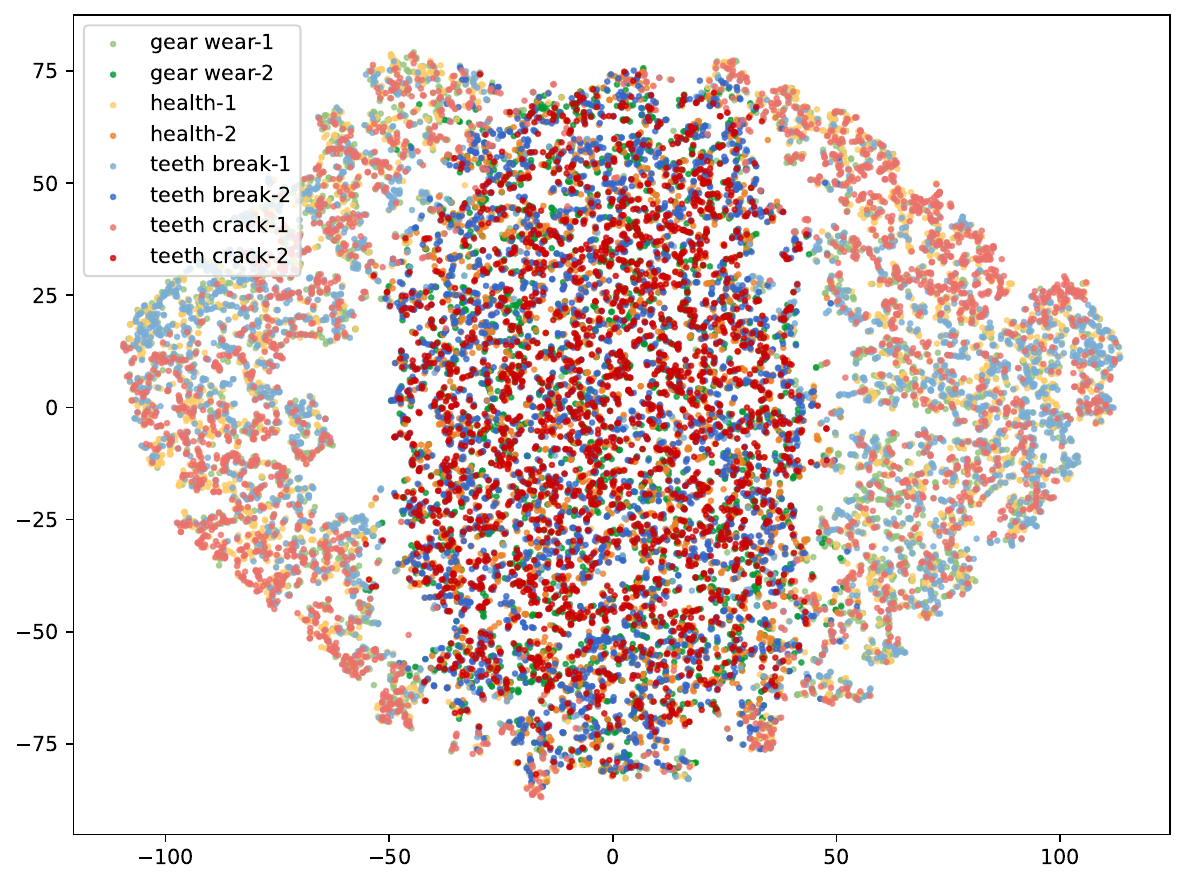}
        \label{raw_speed}
	}
	\centering
	\subfigure[Time-varying speed, extracted features]{
		\includegraphics[width=0.22\textwidth]{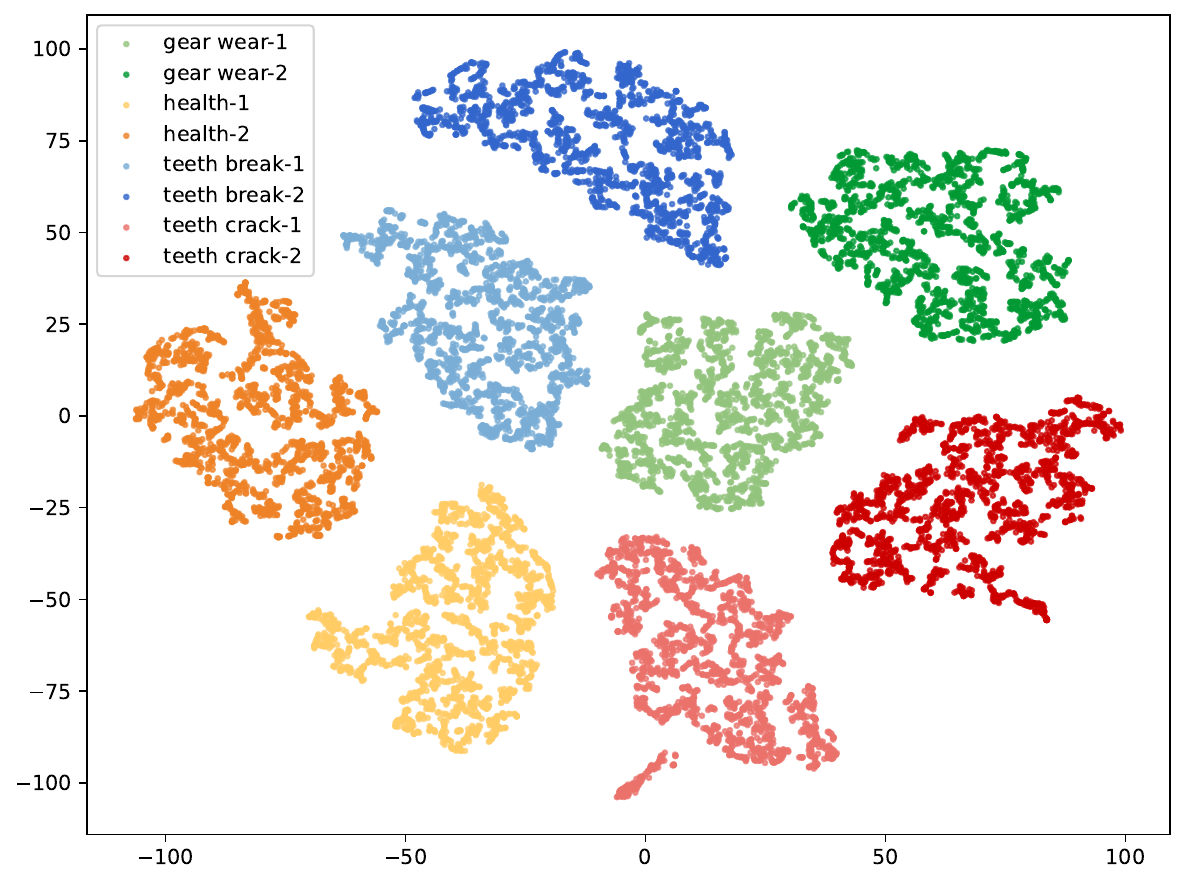}
        \label{mapped_speed}
	}\\
	\centering
	\subfigure[Time-varying torque, raw signals]{
		\includegraphics[width=0.22\textwidth]{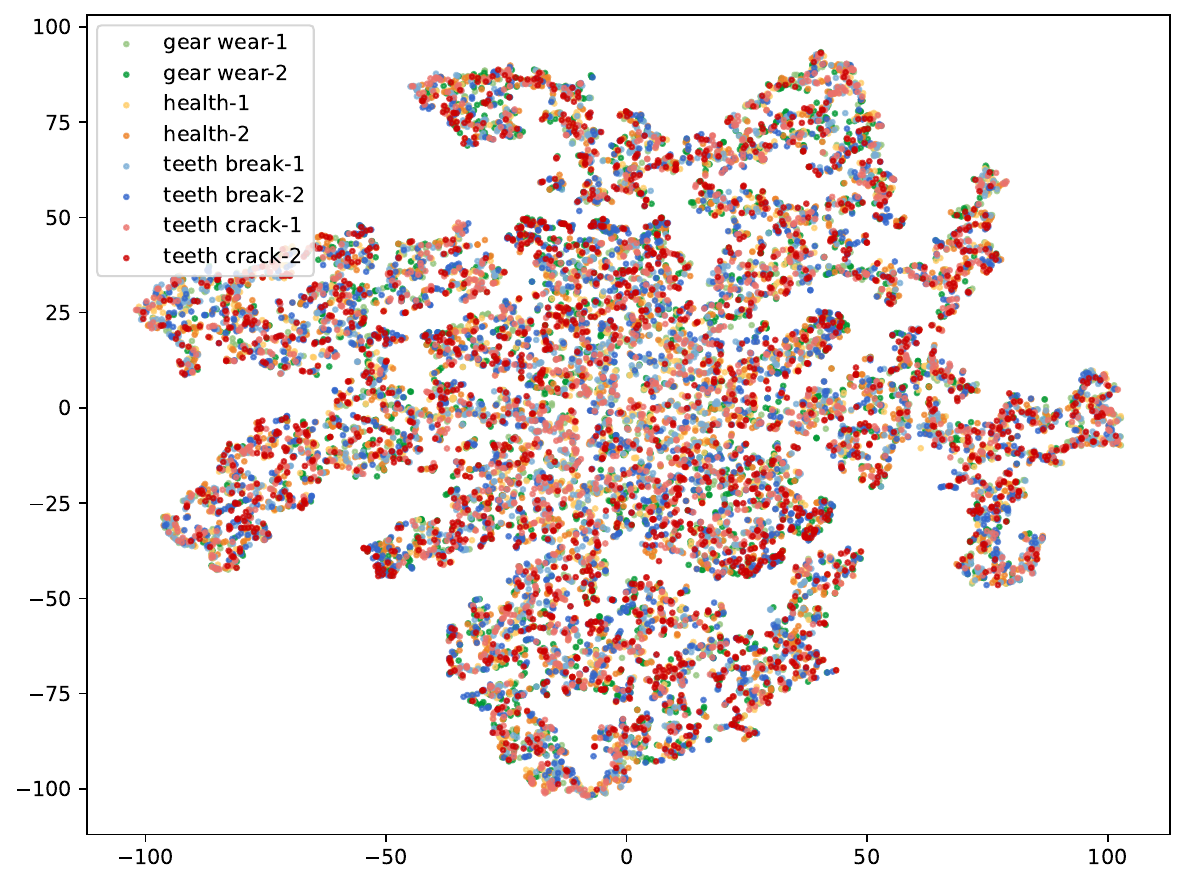}
        \label{raw_torque}
	}
		\centering
	\subfigure[Time-varying torque, extracted features]{
		\includegraphics[width=0.22\textwidth]{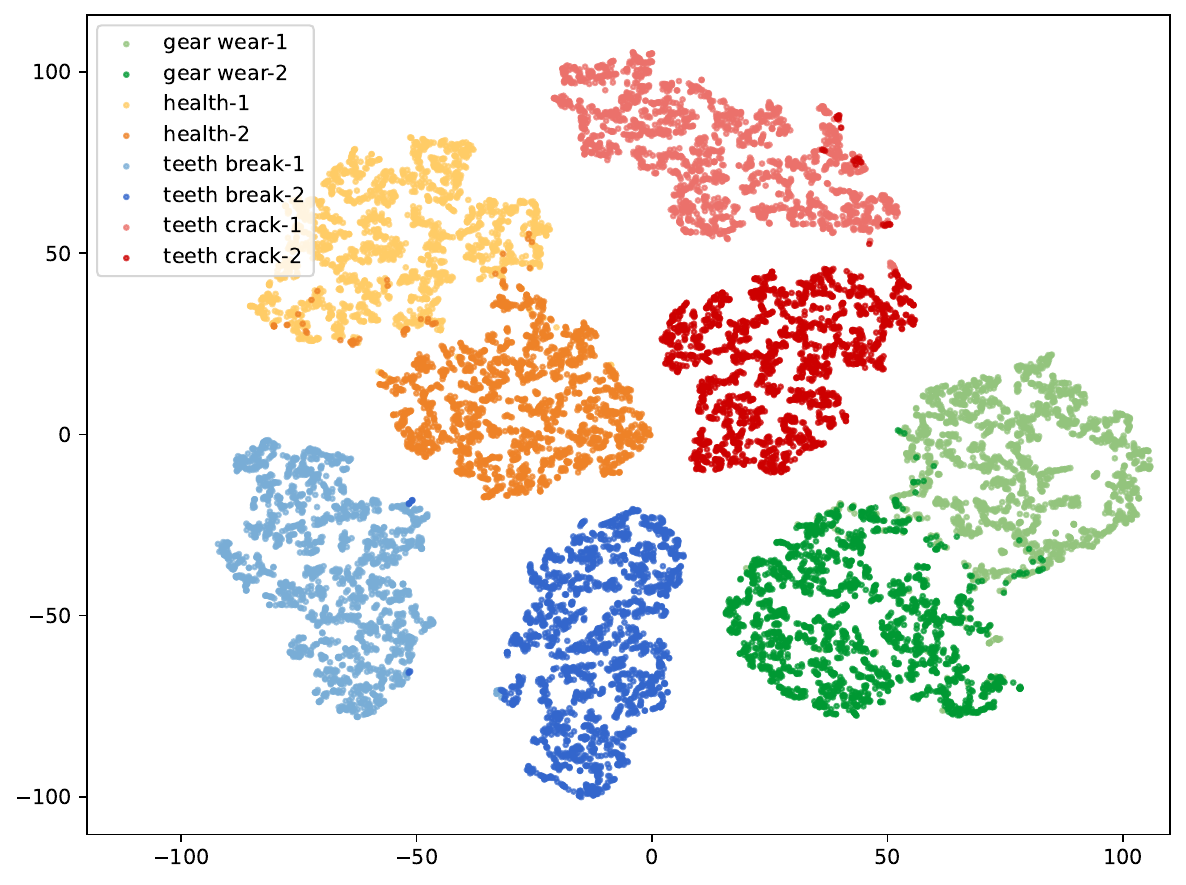}
        \label{mapped_torque}
	}
	\caption{Visualization of dimensionality-reduced features from the offline training set.}
	\label{offline_viz}
\end{figure}

For the raw vibration signals under variable-speed conditions, clear separation between different operating conditions can be observed. In contrast, samples from different conditions in the variable-torque scenario are not easily distinguishable.  
This further confirms that variations in rotational speed introduce more significant condition-related differences than changes in torque.  
Moreover, the t-SNE results show that the samples mapped by the domain-generalized encoder can be easily separated by both condition and fault type, indicating the effectiveness of the feature extraction process.

\section{Conclusion}

\color{black}
In practical industrial processes, varying operating conditions commonly occur, posing significant challenges to traditional fault diagnosis methods.
Although numerous methods have explored the use of transfer learning and other techniques for multi-condition fault diagnosis in recent years, achieving good performance in some scenarios. However, research on the extent to which operating condition information affects fault features has remained limited.
In this context, this study has explored the relationship between operating conditions and faults.
In environments with small condition differences, domain generalization methods based on deep learning have been shown to effectively learn fault features that are invariant to operating conditions, achieving accurate fault diagnosis in varying conditions.
However, in environments with large condition differences, operating condition information can obscure fault-related information.
As a result, the model may have learned domain-specific information during training, leading to overfitting and a decline in overall generalization ability.
To mitigate the negative impact of operating condition information on the model's generalization ability, this paper has proposed a two-stage diagnostic framework.
The domain-generalized encoder is first trained using data from multiple steady operating conditions via domain generalization methods.
Subsequently, a shallow classifier is retrained using the features extracted by the DGE.
By employing the retraining strategy, the coupling between the model and operating condition information can be alleviated, thereby improving fault diagnosis performance.
Several experiments on a real gearbox dataset have demonstrated the effectiveness of the proposed framework compared to the end-to-end model.

\bibliographystyle{ieeetr}
\bibliography{MCFD_250406}

\end{document}